\documentclass[pre, twocolumn, showpacs,amsfonts,amsmath,amssymb]{revtex4}

\usepackage{graphicx}
\usepackage{bm}

\begin{document}

\preprint{APS/123-QED}

\title{Simulated Dark-Matter Halos as a Test of Nonextensive Statistical Mechanics}

\author{Chlo\'{e} F\'{e}ron}
\email{chloe@dark-cosmology.dk}
\author{Jens Hjorth}

\affiliation{Dark Cosmology Centre, Niels Bohr Institute, University of Copenhagen\\ Juliane Maries Vej 30, DK-2100 Copenhagen, Denmark}

\date{\today}

\begin{abstract}
In the framework of nonextensive statistical mechanics, the equilibrium
structures of astrophysical self-gravitating systems are stellar polytropes,
parameterized by the polytropic index $n$. By careful comparison to the
structures of simulated dark-matter halos we find that the density profiles,
as well as other fundamental properties, of stellar polytropes are
inconsistent with simulations \textit{for any value of $n$}. This result
suggests the need to reconsider the applicability of nonextensive statistical
mechanics (in its simplest form) to equilibrium self-gravitating systems.
\end{abstract}

\maketitle

\section{Introduction}

Nonextensive statistical mechanics is a generalization 
of thermodynamics and statistical mechanics proposed by C.\ Tsallis in 1988 
\cite{tsallis88}, aiming to overcome the limitations of the Boltzmann entropy 
in its conventional applications \cite{karlin02,taruya03b,chavanis05,gross05}. 
The theory has considerably widened the fields of 
application of statistical physics, allowing the description of systems 
affected by nonlocal effects, such as long-range forces and memory effects. 

The study of astrophysical self-gravitating systems was one of the first 
applications of the theory \cite{plastino93}. This approach has recently
witnessed a renewed interest in the hope that it may provide a theoretical 
basis for the description of the universal structure of dark-matter (DM) 
halos \cite{hansen05,leubner05,kronberger06,zavala06}. 
Nonextensive statistical mechanics predicts their equilibrium states to be 
stellar polytropes (SP) \cite{plastino93,taruya03}, which have been
claimed  to fit DM halos as well as the usual 
Navarro, Frenk \& White model \cite{kronberger06,zavala06} (hereafter NFW \cite{NFW96}). Moreover, SP have the distinct advantage of being analytically derived from 
nonextensive statistical mechanics, while most models describing DM halos are 
empirical fits to N-body simulations. In this context, nonextensive 
statistical mechanics appears as an attractive framework for providing a theoretical understanding of the structure of DM halos and self-gravitating 
systems in general. 

In this paper we compare, in a parameter independent way, the equilibrium configuration of astrophysical self-gravitating systems predicted by nonextensive statistical mechanics to simulated DM halos. We clarify the issue of which central boundary conditions to use for comparing SP to simulated halos.  On this basis, we establish that simulated DM halos do not corroborate the predictions of nonextensive statistical mechanics. This result calls into question the direct applicability of nonextensive statistical mechanics to equilibrium astrophysical self-gravitating systems.

\section{Nonextensive statistical mechanics and stellar polytropes}

Nonextensive statistical mechanics is based on the formulation of a generalized entropy 
\begin{equation}\label{eqn:Sq}
S_q=k_B \frac{1-\sum_{i=1}^W p_i^q}{q-1} ~(q\in R), 
\end{equation}
which reduces in the limit $q\rightarrow 1$ to the Boltzmann-Gibbs entropy 
$S_{BG}=-k_B \sum_{i=1}^W p_i \ln p_i$
 ($p_i$ is the probability of finding the system in the microstate $i$, $W$ is the number of microstates $i$ for a given macrostate, and $k_B$ is the Boltzmann constant). When the index $q\neq 1$, the entropy of the system is nonextensive, i.e., it is not possible to sum the entropies $S_q(A)$  and $S_q(B)$ of two subsystems $A$ and $B$, but the generalized entropy $S_q$ is conveniently pseudo-additive: $S_q(A+B)=S_q(A)+S_q(B)+(1-q)S_q(A)S_q(B)$. The third term accounts for coupling between the subsystems, due to nonlocal effects.

In this framework, astrophysical self-gravitating systems at equilibrium are described by SP.  This solution arises from extremizing the entropy $S_q$  under the constraints of fixed total mass and energy as derived in \cite{plastino93,taruya03} (see Ref. \cite{tsallis98} about the choice of constraints), leading to a spherically symmetric isotropic system. The resulting distribution function depends on the parameter $q$ and, via the identification $n=1/2+\frac{1}{1-q}$ \footnote[1]{An earlier version of the theory used $n=3/2+\frac{1}{q-1}
$, where the relation between $n$ and $q$ depends on the choice of the constraints when extremizing the generalized entropy $S_q$ \cite{tsallis98,taruya03}. A third relation $n=\frac{5-3q}{2(1-q)}$ has been proposed in \cite{lima05}.}, has the same form as the SP distribution function: $f(\epsilon)\propto\epsilon^{n-3/2}$ for $\epsilon>0$ and $f(\epsilon)=0$ for $\epsilon<0$, where $n$ is the polytropic index and $\epsilon$ the relative energy per unit mass \cite{binney87}.

\begin{figure}
\includegraphics[scale=0.40]{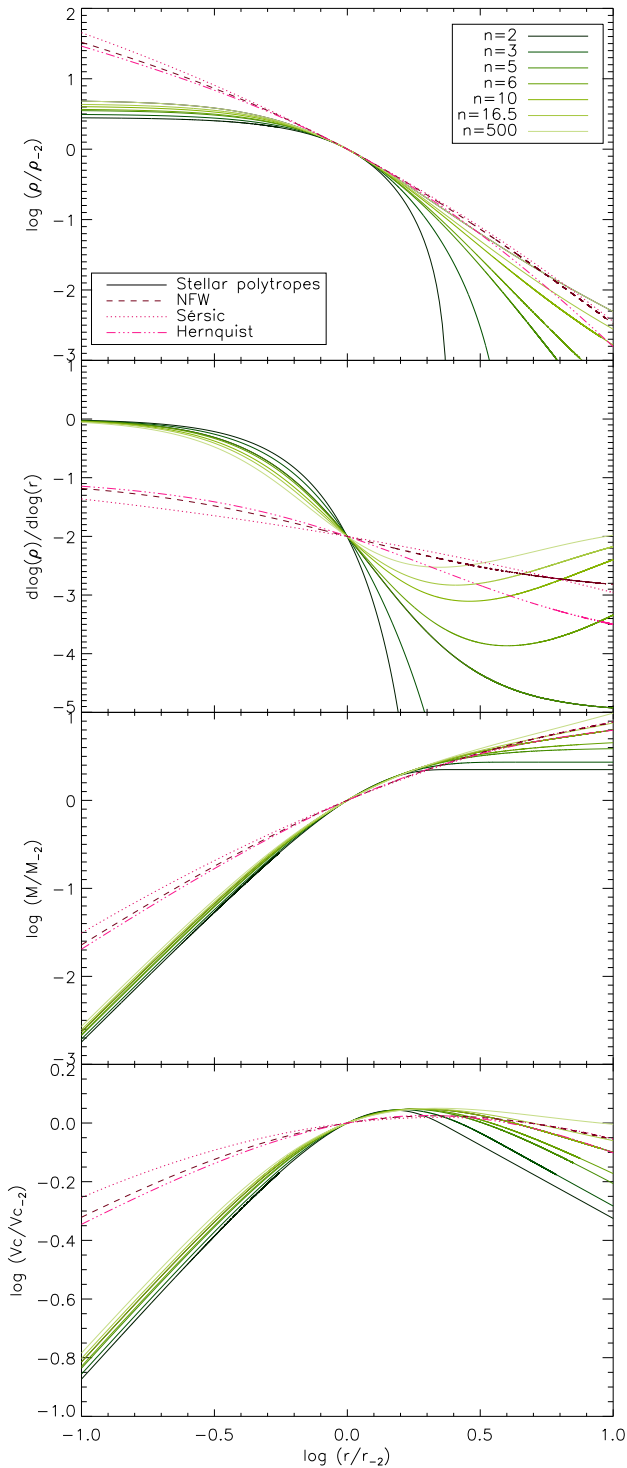}
\caption{\label{fig:figure1} From top to bottom are presented the density, the logarithmic density slope, the integrated mass and the circular velocity. The profiles are scaled to $r_{-2}$, the radius at which the slope equals $-2$. Stellar polytropes predicted by nonextensive statistical mechanics are represented by the solid curves with the shades of green corresponding to different polytropic index $n$.  Simulated dark-matter halos are represented by the NFW model (pink dashed curve), the 3D S\'ersic model (pink dotted curve; $\alpha=0.17$) and the Hernquist model (pink dash-dotted curve; $a=0.45$).}
\end{figure}

In order to compare the predictions of nonextensive statistical mechanics to the results of N-body simulations, we compute radial profiles of fundamental quantities of astrophysical self-gravitating systems:  the matter density $\rho(r)$, the logarithmic density slope $\mathrm{d} \log(\rho)/ \mathrm{d} \log(r)$, the integrated mass $M(r)$, and the circular velocity $V_c(r)$ (see Fig.~1). SP are solutions of the Lane-Emden equation \cite{binney87}, depending on three free parameters: the polytropic index $n$, the central density $\rho_0$ and the central velocity dispersion $\sigma_0$. The value of $n$ (and $q$) is an intrinsic property of each system, reflecting its degree of nonextensivity, but has not been determined from first principles and is therefore a free parameter. SP with $n>3/2$ are stable due to Antonov's stability criterion, i.e., $\mathrm{d} f(\epsilon)/\mathrm{d} \epsilon < 0$ \cite{binney87}. Other values of $n$ are unrealistic: $n=3/2$ corresponds to a distribution function independent of $\epsilon$, and $n<3/2$ to a distribution function diverging at the escape energy, $\epsilon=0$. Therefore, we compute numerical solutions of the Lane-Emden equation for $n>3/2$, choosing values of $n$ ranging from $2$ to $\infty$. We verified the consistency of our profiles using the analytical solutions of the Lane-Emden equation, as well as the isothermal sphere which is the asymptotic solution when $n\rightarrow\infty$ \footnote{Analytical solutions exist only for $n=0,1,5$. We approximate $n\rightarrow\infty$ by $n=500$, as large values of $n$ tend asymptotically towards the isothermal sphere.}.

Nonextensive statistical mechanics does not specify the boundary conditions for SP, which are needed to solve numerically the Lane-Emden equation. Simulated halos necessarily have a finite gravitational potential at the center, therefore we look for SP fulfilling this condition. For SP the density $\rho$ is linked to the gravitational potential $\psi$ by $\rho= cst \times \psi^n$ \cite{binney87}, so we require solutions of the Lane-Emden equation with finite density at the center. Chandrasekhar proved that such solutions have $\mathrm{d} \log(\rho)/\mathrm{d} \log(r)=0$ at $r=0$ \cite{chandrasekhar39}. Therefore, the class of SP we can compare consistently to simulated halos have the natural boundary conditions at the center: $\rho=\rho_0$ and $\mathrm{d} \log(\rho)/\mathrm{d} \log(r)=0$. We present the corresponding profiles in Fig.\ 1. In \cite{kronberger06, barnes07}, the authors arbitrarily fix a non-zero logarithmic density slope near the center, chosen to match the NFW density profile \footnote{E.\ I.\ Barnes, private communication; E.\ van Kampen, private communication.}. This class of  solutions must be treated with caution, as it does not ensure physical boundary conditions at the center.

\section{Comparison to simulated  DM halos}

Cold dark-matter (CDM) halos resulting from N-body simulations provide an excellent way of testing the predictions of nonextensive statistical mechanics. They are collisionless systems whose particles interact via long-range gravitational interaction only, allowing a direct comparison to theory. DM structures in N-body simulations evolve from primordial fluctuations via hierarchical clustering to form halos of universal shape, well described by the NFW model \cite{NFW96}. 
These halos formed through accretion and mergers are sufficiently similar to those obtained by monolithic collapse, i.e., isolated halos, so that we do not take into account in this study the influence of formation history and cosmological environment on the final state of simulated halos \cite{Huss99}.

Collisionless systems have very long collisional relaxation timescales to reach the 'true' equilibrium, but access a stable quasi-equilibrium state faster (dynamical timescales are short compared to cosmological timescales, i.e., the age of the universe) through  the processes of violent relaxation and  phase-mixing \cite{binney87}. 
The quasi-equilibria predicted by numerical cosmological simulations and by nonextensive statistical mechanics, respectively, may in either case be considered the most probable state the system reaches.

Many empirical models have been proposed as fitting the universal profile of CDM halos. We use in our study three of them to represent simulated halos. The NFW model is the more commonly used \cite{NFW96}. The S\'ersic model has been proposed recently as fitting more accurately the inner part of high resolution halos \cite{navarro04, graham06}. The Hernquist model provides a simple description of self-gravitating systems, for which all the profiles we use have analytical expressions \cite{hernquist90}. As seen in Fig.~1, these profiles form a group with similar shapes, and we do not need to use actual simulations in our study.

To compare simulated DM halos to SP, we select the radial range for which recent N-body simulations are robustly resolved. Based on the high resolution simulations published in Navarro et al.\ \cite{navarro04}, this corresponds to $0.1r_{-2}<r<10r_{-2}$, where $r_{-2}$ is the radius at which the logarithmic density slope equals $-2$. To compare SP of all values of $n$ with simulated halos, we need a scaling suitable for finite and infinite mass halos, circular velocity profiles with and without maximum, and density profiles with very different steepness: $r_{-2}$ provides such a universal scaling \footnote{The scaling radius $r_{-2}$ is determined uniquely in the case of a monotonically decreasing slope. However, SP of $n~\geq~5$ have a logarithmic density slope oscillating around $-2$ \cite{medvedev01, binney87}; we therefore define $r_{-2}$ as the smallest radius at which the slope is $-2$.}. Moreover, it allows to take  equally into account the inner and outer parts of the halo, respectively defined as having a slope shallower and steeper than $-2$ \cite{NFW96}. Finally, our study is independent of scaling parameters: SP profiles are scaled depending on the choice of $\rho_0$ and $\sigma_0$, but the shape depends only on $n$. Therefore, scaling to $r_{-2}$ makes identical any profiles of same $n$ but different $\rho_0, \sigma_0$, keeping only information about the shape. All the profiles we present, as scaled to $r_{-2}$,  depend only on structural parameters: range of $n$ for SP; $\alpha=0.17$ for the 3D S\'ersic profile \cite{navarro04}; $a=0.45$ for the Hernquist profile \cite{hernquist90}; $r_s$ for the NFW profile \footnote[2]{In the NFW model, the scale radius $r_s$ is equal to $r_{-2}$.}. In Fig.~1 we compare SP with simulated CDM halos, scaling to $r_{-2}$.

In the outer parts, where $r>r_{-2}$, SP have very different shapes depending on the polytropic index $n$. Profiles with $n<5$ correspond to finite-mass halos, while $n\geq5$ polytropes have infinite mass, tending to the isothermal sphere when $n\rightarrow\infty$. While finite-mass polytropes may appear more attractive from a physical point of view, they also provide worse fits to simulated DM halos, with an outer slope as steep as $\mathrm{d} \log(\rho)/\mathrm{d} \log(r)<-5$ at $10r_{-2}$. 
For infinite-mass SP with larger values of the polytropic index, $n=16.5$ (as found in \cite{kronberger06}) agrees with NFW and S\'ersic models, $n=10$ with the Hernquist model. Therefore, \textit{some infinite-mass SP can provide a good description of the outer parts of simulated halos}. 

In the inner parts, where $r<r_{-2}$, SP share similar properties: they have a large core, the shape and the extent of which depend little on the value of $n$. This feature is in striking disagreement with the predictions of N-body simulations, which show steeper inner slopes.
While this core structure itself has been advanced as an advantage of SP over the NFW profile \cite{zavala06}, solving the well-known cusp problem between simulations and observations \cite{gentile04}, it appears in Fig.~1 that this is not the case: the difference between SP and simulated halos in the density, density slope and mass profiles is as high as one order of magnitude at $0.1r_{-2}$.  Therefore, \textit{for any polytropic index $n$, SP do not properly describe the inner parts of simulated halos}.

\section{Discussion}

We have established here that the predictions of nonextensive statistical 
mechanics are not corroborated by simulations of DM halos. Our results are 
based on readily observable quantities, i.e., the matter density itself, and  
profiles derived from it. These findings are in direct contrast to previous 
works which found reasonable agreement because they either considered only 
the outer parts of simulated halos \cite{leubner05}, or used a non-zero density slope as initial condition near the center \cite{kronberger06}, or used 
values of $n<5$ that lead to too steep density profiles in the outer parts \cite{zavala06}. 
Support for our conclusions, however, comes from Barnes et al.\ \cite{barnes07} who, even though fixing a non-zero density slope near the center to match the NFW density profile, found inconsistency between SP and DM halo velocity dispersion profiles. While we are cautious about the physical relevance of such profiles, it is of interest to emphasize that nor cored neither cuspy polytropes can describe properly simulated DM halos.

These results imply a fundamental difference in the matter distribution of 
astrophysical self-gravitating systems predicted by nonextensive statistical 
mechanics and by N-body simulations.
Such a discrepancy raises three main questions. Is our comparison with  simulated DM halos valid? Is nonextensive statistical mechanics the proper theory to 
describe collisionless long-range interaction systems at equilibrium? 
And how does this study relate to observed DM halos?
We briefly address these issues below.

1a. We study idealized systems which do not take into account complex effects in DM halos such as velocity anisotropy \cite{hansen06c} or triaxiality \cite{Hayashi07}. However, the error we make by using isotropic and spherically averaged models to represent DM halos is small compared to the discrepancy we find between SP and simulated halos. Moreover, we do not address here general problems of statistical mechanics of self-gravitating systems like, e.g., infinite mass \cite{hjorth91}, so as to focus on the issues specific to nonextensive statistical mechanics.

1b. N-body simulations depend on various parameters, such as the choice of the softening length and the number of particles, which can introduce numerical effects in the resulting halos. The softening of the gravitational force on small scale, introduced to avoid two-body interaction, creates a core at the center of the halo, approximately the size of the softening length. The number of particles fixes the maximum phase-space density resolved at redshift zero, and if not large enough, a core appears due to the lack of particles in the center. Both effects lead to a shallower density profile in the center of halos if the simulation is of low resolution \cite{moore98}. Therefore, the discrepancy we observe between simulated profiles and SP is genuine and increases with the resolution of numerical simulations.

1c. We compare theoretical statistical predictions to cosmological simulations, but the latter depend on initial conditions and cosmological parameters. Initial conditions are fixed by the shape of the power spectrum of initial fluctuations $P(k) \propto k^{n_s}$, i.e., by the choice of the index $n_s$. From CMB observations, we have an indication that  $n_s=0.958 \pm 0.016$ \cite{spergel07}, but tests on cosmological simulations proved that they depend very little on the choice of $n_s$ and on the cosmological model used \cite{navarro97}. Therefore, the universal profiles of simulated halos can be compared to statistical theories as a general prediction.

1d. The choice of a scaling is necessary to compare DM halos models, but can bring a visualization bias. We checked if the use of $r_{-2}$ leads  us to overestimate the disagreement between nonextensive statistical mechanics and N-body simulations. However, adjusting the fit in the outer parts of the halo, e.g., scaling to the virial radius, increases even more the discrepancy in the inner parts, while scaling profiles to fit well at smaller radii makes the discrepancy appear in the outer parts of the halo. 

2. Nonextensive statistical mechanics relies on three assumptions: \textit{a)} the generalized entropy $S_q$ is the right form to describe long-range interaction systems; \textit{b)} the system is at equilibrium, and its most probable state is given by the maximum entropy principle; \textit{c)} $S_q$ is maximized at fixed energy, leading to a system whose distribution function depends only on the energy per unit mass, and has isotropic velocity dispersion. 
Though simulated halos show evidence of velocity anisotropy, the relation between density slope and velocity anisotropy \cite{hansen06c} shows that it is isotropic in the center, where violent relaxation \cite{lyndenbell67} (the process by which
strong potential fluctuations efficiently drives
the system towards equilibrium) is most effective.
Therefore, the assumptions \textit{b)} and \textit{c)} hold true in the center.   However, it is in the inner parts that the disagreement between SP and simulated halos is the strongest. Hence, we suggest that the assumption \textit{a)}, i.e., the choice of the
generalized entropy $S_q$ motivated by nonextensive statistical mechanics,
cannot be used to predict the equilibrium structures of simulated DM halos.

3. Simulated DM halos are in good agreement with observations \cite{pointecouteau05}, except for the inner core found from spiral galaxy rotation curves \cite{gentile04}. Stellar polytropes, interestingly,  have an inner core too, but the discrepancy we observe between SP and simulated DM halos is too large to explain the core of observed DM halos. Adding adiabatic contraction in simulations results in DM halos with steeper inner parts, which would not change our conclusions.

In summary, we have established that nonextensive statistical mechanics 
\cite{tsallis88, plastino93, taruya03}, a theory generalizing classical statistical mechanics and thermodynamics, does not describe the equilibrium state of astrophysical self-gravitating systems, as represented by DM halos formed in N-body simulations.

\begin{acknowledgments}
It is a pleasure to thank S.\ H.\ Hansen, E.\ van Kampen and E.\ I.\ Barnes for valuable discussions and comments, and to acknowledge the referees for helping us to improve this paper.
The Dark Cosmology Centre is funded by the Danish National Research Foundation.
This work was supported in part by the European Community's Sixth Framework Marie Curie research Training Network Programme, Contract No.\ MRTN-CT-2004-505183 
``ANGLES".
\end{acknowledgments}


\end{document}